\documentstyle[12pt]{article}
\newcommand{\be}{ \begin{eqnarray}}
\newcommand{\ee}{\end{eqnarray}}
\newcommand{\beno}{ \begin{eqnarray*}}
\newcommand{ \eeno}{\end{eqnarray*}}
\newcommand{\raf}[1]{(\ref{#1})}
\begin{document}
\bibliographystyle{try}
 \begin{titlepage}
 \begin{center}
\ \\
{\Large \bf Effects of Collective Potentials on Pion Spectra
in Relativistic Heavy Ion Collisions}
\vspace{2cm}
\ \\
{\large V. Koch
\footnote{Supported in part by Deutsche
Forschungsgemeinschaft and in part by the U.S. Dept. of Energy
Grant No. DE-FG02-88ER40388},}
\ \\
{\it Physics Department, State University of New York\\
Stony Brook, NY 11794, U.S.A.}\\

\ \\
{\large G.F. Bertsch
\footnote{Supported the National Science Foundation under
Grant No. 90-17077}}
\ \\
{\it Physics Department and NSCL, Michigan State University,\\
East Lansing, MI 48824}\\

\vspace{2cm}
{\large \bf Abstract}\\
\vspace{0.2cm}
\end{center}
 \begin{quotation}
The effect of collective potentials on pion spectra in ultrarelativistic
heavy ion collisions is investigated. We find the effect of these potential to
be very small, too small to explain the observed enhancement at low
transverse momenta.
 \end{quotation}
 \end{titlepage}
\newpage
\centerline{1. Introduction}
\ \\

The enhancement at low transverse momentum found in negative particle
and neutral pion spectra from ultra relativistic heavy--ion collisions
\cite{Str89,Ake90,Alb90} has recently received considerable
interest in the literature. One explanation for
instance \cite{LH89,AFK87} has interpreted the enhancement as a
collective flow effect, but more detailed considerations concerning
the freeze out surface result in too small an effect \cite{KB89}. The
decay of excited baryons also gives rise to additional soft
pions \cite{BSW91,BBK91,SKH90}. However, at CERN-energies there are
simply not enough baryons present in the central region in order to
account for the observed enhancement \cite{BBK91}. Furthermore the
combined spectrum of thermal plus decay pions still flattens out at
very low p$_t$ contrary to the data.

Kataja and Ruuskanen \cite{KR90} have shown that the measured spectrum
can be fitted by assuming that the pions are strongly out of
equilibrium.
Although one may give some qualitative arguments \cite{GLG90}
why an excess of pions should build up during the expansion,
currently there is no quantitative understanding on how a chemical potential
of the size needed to fit the data ($\mu \simeq
130~MeV$) would arise.

In refs. \cite{WB91} and \cite{BBD92} a kinetic model
with Bose-statistics in the collision integral has been used to study
the build up of the low p$_t$ enhancement assuming different initial conditions
for the expanding source. The authors find that the Bose-statistics in the
collision integral indeed leads to an enhancement provided the initial pion
density is sufficiently high. They also point out that corrections of the
cross section due to Bose statistics are large and that they
limit the effect of the
Bose-phase space factors in the collision integral substantially.
Using the corrected cross section and allowing for higher resonances in the
initial state the authors have to assume a hadronization
time as short as $\tau_0 \sim 1fm$ in order
to account for the measured enhancement by the hadronic scattering processes
only.

Finally the pions may interact collectively with the surrounding hadronic
medium. As proposed by Shuryak \cite{Shu91,Shu90} these interactions
may give rise to a strongly momentum dependent optical
potential which is
attractive for low momenta and, therefore, could lead
to an enhanced soft component in the pion spectrum.

It is the purpose of this article to study the effect of this latter mechanism
in detail. In the first part we will
develop the mean field potential and demonstrate how it can be used in a
transport theoretical framework. Then we will study the effect of the mean
fields for a static potential. Finally the expansion of the fireball will be
taken in account. This will be done using a
transport model which combines the propagation
of particles in the mean field as well as the collisions among the particles.

\newpage
\centerline{\bf 2. Mean Field Potential}
\ \\

Following Shuryak \cite{Shu91},
the most important contribution to the collective
potential felt by a pion is the coherent
scattering of two pions via the P-wave $\rho$--resonance; in the S-wave
the contributions from different isospin channels cancel each other.
We first show that these collective potentials can be
derived from an
energy functional by differentiating with respect to the distribution function.
As a consequence energy conservation will be guaranteed and the collective
potentials can be incorporated similar to the nuclear mean field in transport
models of heavy ion collisions \cite{BD88,BKM92}.

For an arbitrary resonance the energy functional has the following form:
\be
H = g_{I,J} \int \frac{d^3p}{(2 \pi)^3 2 \omega (p)}
\frac{d^3p'}{(2 \pi)^3 2 \omega (p')} \sum_l f_l(x,p) \, \sum_m f_m(x,p')
\frac{4 \pi \sqrt{s}}{q}
\frac{(\sqrt{s} - M) \, \Gamma (q) }{(\sqrt{s} - M)^2 + \Gamma^2(q) / 4}
\ee
with q and $\sqrt{s}$ being the c.m. momentum and energy and
$M$ and $\Gamma$ are the mass and the width of the resonance under
consideration
e.g. the $\rho$.
The phase-space distribution $f_l(x,p)$ contain an index which refers to the
internal quantum numbers such as spin and isospin. These quantum numbers are
summed over. The degeneracy factor
$g_{I,J}$  is given by
\be
g_{I,J} = \frac{1}{1 + \delta_{p1,p2}} \,
\frac{(2J_{res}+1)(2I_{res} + 1)}{(2J_{p1}+1)(2I_{p1} + 1) \,
(2J_{p2}+1)(2I_{p2} + 1)}
\ee
where the indices $p1$ and $p2$ refer to the particles forming the resonance.
For example in the case $\pi+\pi \rightarrow \rho$ we would have
$g_{I,J} = 1/2$

The resulting mean field potential for a particle of given internal quantum
number $l$ is then given by
\be
U_l(p,x) = \frac{\delta H}{\delta f_l} =
g_{I,J} \frac{1 + \delta_{p1,p2}}{2 \omega (p)} \int
\frac{d^3p'}{(2 \pi)^3 2 \omega (p')} \sum_m f_m(x,p')
\frac{4 \pi \sqrt{s}}{q}
\frac{(\sqrt{s} - M) \, \Gamma (q) }{(\sqrt{s} - M)^2 + \Gamma^2(q) / 4}
\label{eq:2}
\ee
which in the special case of $\pi+\pi \rightarrow \rho$ leads to
the result already obtained by Shuryak. Assuming isospin symmetry we obtain:
\be
U_{\pi}(p) = \frac{3}{2 \omega (p)} \int \,
\frac{d^3p'}{(2 \pi)^3 2 \omega (p')} \bar{f}(x,p')
\frac{4 \pi \sqrt{s}}{q}
\frac{(\sqrt{s} - M_{\rho}) \, \Gamma_{\rho} (q) }{(\sqrt{s} - M_{\rho})^2
+ \Gamma_{\rho}^2(q) / 4}
\label{eq:2b}
\ee
where $\bar{f}(x,p) = 1/3 \sum_{l=+,-,0} f_l(x,p)$
stands for the isospin averaged pion phase-space distribution.

In the following we will truncate the explicit energy dependence
of the potential
$U$ by evaluating \raf{eq:2} on the mass shell only, i.e. $\omega^2 = p^2 +
m_\pi^2$, which is equivalent to taking only the first contribution in the
Dyson series for the self energy.
As a consequence the potential depends on the 3--momentum of the particle
only  and, therefore, the pion wave function will not be modified.
Hence we can
treat the mesons as quasiparticles and a transport theoretical approach is
possible\footnote{Note that the formalism described here may used as well for
pions interacting with nucleons to study in--medium effects on pions
in lower energy collisions.}.

For local equilibrium the distribution function can be written  as

\be
f(x,p) = \rho_0 \exp(- E / T(x) )
\label{eq:3}
\ee
where $E = \sqrt{p^2 + m^2}$ denotes the energy and
$T(x)$ the local temperature.
Using the above phase--space distribution \raf{eq:3} we can calculate the
mean--field potential $U(p)$ from eqns. \raf{eq:2} or \raf{eq:2b}.

In figure \ref{fig:pot2}
the resulting pion collective potential due to
$\pi+\pi \rightarrow \rho$ are shown for
three different temperatures
($T = 150, \, 200, \, 250 \, \rm MeV$) (the full lines).
We find the potential  becomes deeper with increasing temperature,
because at the same time the density of
pions in the heat-bath increases.
The point where the potential changes sign, on
the other hand, is essentially unaffected by the temperature.

Since we are interested in a more general discussion of
mean field effects in this article,
in fig \ref{fig:pot1} we show the dependence of the
collective pion potential on the resonance mass.
These potentials have been obtained for the same
temperature $T= 200 \, \rm MeV$ but with different resonance mass $M_{res} =
500,\, 770,\, 1100 \, \rm MeV$.
For the width we have taken a p-wave parameterization with a value of
$\Gamma_0 = 150 \, MeV$ on resonance and the degeneracy factor was chosen to be
the same as in $\pi+\pi \rightarrow \rho$.
As we would have expected from eqn. \raf{eq:2}
the potentials changes sign at a momentum close to the resonance mass.

Since in the transport theoretical calculation described below
the mean field is evaluated assuming a thermal momentum distribution,
it is useful to parameterize the above potential in a simple
form
\be
U(p) = V_0 \, (1 - (\frac{p}{a_1})^2) \, \, \exp(-(p/a_1)^2) \,
(\frac{T}{T_0})^3
\label{eq:4}
\ee
where $V_0$ is the value of the potential at zero momentum while $a_1$
corresponds to the momentum where the potential changes sign.
The temperature dependence in this parameterization reflects the fact that the
optical potential is essentially proportional to the pion density.
In figure \ref{fig:pot2} we compare this parameterization for three different
temperatures ($T = 150, 200, 250 \, \rm MeV$) with the potential based on
$\pi+\pi \rightarrow \rho $. The parameters for the fit are
$V_0 = -.2 \, m_\pi$, $a_1 = 4 m_\pi$ and $T_0 = 200 \, \rm MeV$.
The agreement is reasonable over the whole range of temperatures displayed.
For the sake of simplicity in the following we will only
use the parameterization \raf{eq:4} for the mean field potential. Also the
meaning of the parameters becomes more transparent with this choice.

In following we will use several parametersets for the potential \raf{eq:4}
which we display in table \ref{tab:1}. Set 1 corresponds to the potential
obtained by Shuryak for the pure pion gas. Since our purpose is a general
understanding of mean field effects, we have also calculated with other
parameter sets shown in the table. Both these are much stronger than the
realistic potential, in order to see the effects of a strong potential.

\begin{table}
\centerline{
\begin{tabular}{l||c|c|c}
Parameterset & $V_0 \, [\rm MeV]$ & $a_1 \, [\rm MeV] $ & $ T_0 \, [\rm MeV ]$
\\ \hline
Set 1        & 40      &             650        & 200
\\ \hline
Set 2        & 100     &             650        & 200
\\ \hline
Set 3        & 100     &            1300        & 200
\end{tabular}
}
\caption{Different sets of parameters of effective potential
\protect\raf{eq:4}.}
\label{tab:1}
\end{table}

\newpage
\centerline{\bf 3. Static Potential}
\ \\

Before we turn to the expansion of the fireball it is instructive to first
study
the simpler case of particles leaving a static potential well. Because the
energy $E$ of the particle is conserved while traversing the
potential the
{\em energy} spectra $dN / d E$ inside and outside are identical

\be
\frac{dN}{d E}_{inside} = \frac{dN}{d E}_{outside}
\label{eq:2.1}
\ee

Let us assume that the particles inside the potential are distributed according
to a boost-invariant fire tube

\be
\frac{dN}{dy d^2p_\bot} = \int_{-\infty}^{+\infty} d \eta \, m_\bot \cosh(\eta)
\exp (-\beta m_\bot \cosh \eta ) = 2 m_\bot {\rm K_1} (\beta m_\bot)
\label{eq:2.2}
\ee
where for simplicity we have assumed that the particles are distributed
according
to Boltzmann-statistics ($\beta = 1/T$). Thus, in a given rapidity bin,
the problem reduces to a two dimensional one, with $ E = m_\bot =
\sqrt{m^2 + p^2_\bot}$.

Because of relation \raf{eq:2.1}, the potential only affects the momentum
spectra; for cylindrical symmetry we have:
\be
\frac{dN}{d^2p} = \frac{1}{p} \frac{dN}{dp} = \frac{1}{p} \frac{d E}{dp}
\frac{dN}{d E}
\label{eq:2.3}
\ee
Outside the potential we have
\be
\frac{d E}{dp} = \frac{p}{ E }
\label{eq:2.4}
\ee
thus we find
\be
\frac{dN}{d^2 p_\bot}_{outside} =
\frac{1}{E}  \frac{dN}{d E}_{outside} =
\frac{1}{E}  \frac{dN}{d E}_{inside} =
\frac{1}{E}  \left( \frac{p}{d E / dp} \frac{dN}{d^2 p_\bot}
\right)_{inside}
\label{eq:2.5}
\ee
finally using eqn. \raf{eq:2.2} for the momentum distribution inside the
potential we find
\be
\frac{dN}{d E}_{outside} =
\frac{p_{in}}{(d E / dp)_{in}} \frac{dN}{d^2 p_\bot}_{inside}
\label{eq:2.6}
\ee
where $p_{in}$ is determined from
\be
\sqrt{p_{in}^2 + m^2} + U(p_{in}) = E
\label{eq:2.7}
\ee
and
\be
\frac{d E}{d p_{in}} = \frac{p_{in}}{\sqrt{p_{in}^2 + m^2}} +
\frac{d U}{dp_{in}}
\label{eq:2.8}
\ee

In fig \ref{fig:static} we have plotted the resulting spectra for the different
parameterizations (table \ref{tab:1})
of the potential together with the free one. For the spectra inside the
potential we have used a temperature of $T = 1/\beta = 135 \, \rm MeV$, which
fits the transverse momentum spectra from proton-proton collisions (see below).
All spectra are normalized such that they have the same number of particles
{\em inside} the potential.
Therefore, in fig. \ref{fig:static} the integral over the free spectrum is
larger than the integral over the ones with a potential, because in the latter
case all particles with $E < m$ are bound inside the well.

We find that, as a result of the potential, the spectra become somewhat steeper
at small momenta. The effect, however, is much too small in order to explain
the
data. Even if we increase the depth of the potential well from $V_0 = -40 MeV$
(short--dashed line) to $V_0 = -100 MeV$ (long--dashed line),
the slope of the spectrum is not changed much (aside from an overall shift
downwards due to the fact that the deeper potential keeps more particles bound
inside). The difference between parameterset 2 (short--dashed line) and 3
(long--dashed line) is even smaller.

This behaviour
actually can be understood by looking at figure \ref{fig:disp}, where we have
plotted the resulting dispersion relations $E (p)$
for the three parameterizations. Only particles with energy above the
rest mass $E(p) \geq m$ contribute to the outside spectrum. There the
slopes of the dispersion relation $d E / dp$ do not differ very much
from the free one. As a consequence
the depth of the potential does not affect the spectra very much. Also possible
minima of the dispersion relation \cite{Shu91,Shu90} do not affect the outside
spectrum, because they also occur at energies smaller than the rest mass.

In conclusion, we find, that in the (not very realistic) case of a static
potential well, only small effects of the potential on the particle spectrum
can be observed, too small to account for the observed enhancement.
The main reason is that the particles which may escape the potential
feel only a rather
weak potential with very moderate momentum dependence.

Of course as already pointed out, in reality the
potential is not static because its source, the fireball,
 expands. Consequently the potential decreases as time continues
so that eventually all particles may escape.
The observed enhancement could, therefore, still originate from those
particles, which  would be trapped inside the static potential.

This would be the case if these particles leave the system
as soon as energy conservation allows the to do so, i.e once their energy
becomes larger than their rest mass. However, such a scenario,
depends very much on the time scales involved in the
problem. It would require that the expansion of the fireball is very slow
compared to the velocity of the low momentum particles.
Since the fireball is made out of pions as well this seems to be very unlikely.

On the other hand, if
the fireball expands with a velocity faster than the low momentum particles
these particles will essentially remain inside the potential well (fireball)
until the potential has vanished. In this case only the fast particles would
feel an effect of the potential as they have to climb a potential well of
finite
depth. At the soft part of the spectrum, however, we would not expect
any effects of the potential\footnote{This later scenario has been studied in
great detail in case of anti-protons in ref. \cite{KBK91}.}.

The question of the time scales involved can best be answered in a model
calculation. In the next section we, therefore, will
study the full expansion of the fireball and the effect of the collective
potentials on the spectra in a transport model. This approach should provide a
reasonable simulation of the expansion and the time scales involved.

\newpage

\centerline{\bf 4. Expansion of the fireball}
\ \\

In order to study the dynamic effects of the collective potential
we have extended a cascade model \cite{BB90} to include the propagation
of the particles in a mean field. Following the standard procedure \cite{BD88}
the particle coordinates and momenta are propagated according to Newton's
equations of motion
\be
\frac{dr}{dt} & = & \frac{p}{E} + \frac{dU}{dp}
\nonumber \\
\frac{dp}{dt} & = & -\frac{dU}{dr}
\label{eq:4.1}
\ee
where the mean field potential $U$ is given by eqn. \raf{eq:4}.
The temperature which is needed in order to determine the mean field
potential is
calculated from the density of pions assuming local thermal equilibrium.
The pion-density is extracted from the actual particle distribution.
At every timestep the radial dependence of the density
is fitted with a function of the form
\be
\rho( r_\bot ) = A (1 + b^2 r_\bot^2) \exp(- b^2 r_\bot^2 )
\label{eq:3b}
\ee
The parameters $A$ and $b$ are determined by the root-mean-square radius of
the distribution and the total number of pions.
In the longitudinal direction, on the other hand, for a given timestep the
density
is assumed to be constant between $z_{min}$ and $z_{max}$, where $z_{min}$ and
$z_{max}$ denote maximum distance in positive and negative direction where
particles have materialized. Comparing with the actual density distributions
these assumption are very well justified for the conditions we are dealing
with and which we will discuss below\footnote{This is essentially a result of
the Bjorken initial conditions we have imposed here (see below).}.
The error introduced is certainly not larger than the one one  would have
when using a spatial grid and having to deal with large density fluctuations
\cite{KBC91}.

The cascade includes $\pi$, $\eta$, $\rho$ and $\omega$ mesons but no
baryons\footnote{At CERN energies the ratio of protons to $\pi^-$ is about 1:6
for $S + S$ collisions \cite{Str90}.}.
All mesons decay according to their empirical lifetime and decay channels.
For the pion--pion scattering measured phase
shifts are used while for all other elastic processes a constant cross section
of 20~mb is assumed. The only inelastic process taken into account is
$ \pi + \pi \leftrightarrow \rho $. In addition the $\omega$ is allowed
to decay into three pions.
While this may not take into account
all the details of the meson--meson scattering it certainly provides
enough accuracy in order to lead to a  reasonably realistic expansion
scheme. Finally the
model does not respect the Bose nature of the mesons,
i.e. we do not have any Bose enhancement factors
incorporated in the collision integral
as done e.g. in refs.  \cite{WB91,BBD92}.  Here, we are
rather interested in dynamical effects due to collective
potentials.

We shall specifically be concerned with the central $200
\,GeV/a$ $^{16}$O + Au data of the NA35 collaboration \cite{Str89}.
This experiment measures negative particles and does not
identify the pions explicitly. Usually one assumes a $\sim 10
\%$ admixture of kaons, electrons and anti--protons. For
simplicity, however, we neglect this fact and assume all
negatives to be pions.

The number of initial particles is determined such that the measured
rapidity distribution of pions is reproduced.
We further impose Bjorken--like \cite{B83} initial conditions:
rapidity $y_{boost}$ and
longitudinal coordinate of a locally thermal distribution
are uniquely related by
\be
z = \tau_0 \,sinh \, y_{boost}
\label{eq:5}
\ee
where $\tau_0$ may be considered a formation or hadronization time.
In other words particles materialize, i.e. participate in the expansion, only
after their proper time is larger than the hadronization time $\tau_0$. Because
of eqn. \raf{eq:5} this implies that in the c.m. system particles which are
created at a large longitudinal distance from the center will materialize at a
later time. The initial radial
distribution is assumed to follow the density profile of the
oxygen projectile.

We will study different ways of populating the initial momentum space.
One possibility is to assume
local thermal and chemical equilibrium so that with given density
the initial local temperature and the admixture of higher resonances is fixed.
Thus, in this approach
the free parameters of the model are the initial number of
particles and $\tau_0$.
Those can be determined by fitting the measured
rapidity distribution and slope of the high energy part of the
pion spectrum. The lowest hadronization time which still leads to acceptable
transverse momentum spectra and rapidity distributions is
$\tau \simeq 8 \, \rm fm/c$. The number of initial particles is 380.

The other procedure assumes that the momenta of all
initial particles are distributed
according to a slope parameter for p-p collisions of $T_0 = 135 \, \rm MeV$.
Their relative
multiplicities are distributed proportional to their degeneracy, i.e.
$\pi : \eta : \rho : \omega = 3 : 1 : 9 : 3$
These statistical weights are very
close to what one obtains in string fragmentation models
\cite{AGN87}. In order to reproduce the measured rapidity distribution we
start with 280 particles in the initial state.

The resulting rapidity distributions of pions for both initialization schemes
obtained after 30 fm/c
are shown together with the experimental data of the NA35 collaboration
\cite{Str89} in fig. \ref{fig:rap}.
These results have been obtained with the pure cascade
without potentials. However, since in our model the density and
hence the potentials are assumed to be constant along the
longitudinal direction, the rapidity distribution is not
affected by the potentials.

Let us now turn to the effect of the collective potentials.
In fig. \ref{fig:pt1} and \ref{fig:pt2} we show the resulting transverse
momentum spectra using the different initial conditions described above.
Fig. \ref{fig:pt1} corresponds to the local equilibrium initialization
while fig. \ref{fig:pt2} represents the string fragmentation picture.
Different expansion schemes are studied:
\begin{enumerate}
\item The particles do not interact but may decay (full line).
\item Particles may collide but do not feel any mean field forces
(short--dashed line).
\item Particles collide and are subject to the mean field forces
(long--dashed line and dashed-dotted line).
\end{enumerate}
We display only the results for parametersets 1 and 2 of table \ref{tab:1},
because the results obtained with parameterset 3 are indistinguishable from the
one obtained with parameterset 2.

First let us point out the difference between the expansion with and
without collisions (full and dashed histogram) in the low p$_t$ spectra
exhibited in fig \ref{fig:pt2}. Since the initial conditions do not correspond
to local chemical equilibrium
the collisions are still very effective and provide a
significant enhancement a low transverse momentum. This result should also be
seen in connection with the findings of ref. \cite{BBD92}, where the effect of
Bose statistics in the collision integral is studied. Considering the fact that
our results have been obtained without Bose statistics, we conclude that a
great part of the effect demonstrated in ref. \cite{BBD92} can be explained
without any Bose statistics. This supports the finding pointed out in the
aforementioned reference, namely
that the effect of Bose statistics in the final state
and in the matrix element cancel each other to a large extent.
In case of the local-equilibrium initialization (figure \ref{fig:pt1}) the
effect of the collision is very small. One of the reasons simply is that
we we had to chose a very high hadronization time $\tau_0$, in order to
reproduce the high momentum piece of the spectra. As a consequence the initial
density of particles is comparatively low and thus the collisions are not so
effective.

For either initial conditions the mean field leads only to very little
enhancement at low p$_t$. Also the stronger mean field
parameterization does not provide more enhancement.
Thus, the expansion  does not provide the additional enhancement compared
to the static limit. Quite to the contrary, the
effects found in the static limit are reduced because the time-averaged depth
of
the potential well is smaller as a result of the expansion. And certainly, the
enhancement due to the pions being trapped in the static potential, as
discussed
in the last section, does not show up. Obviously, the velocity of the fireball
expansion is faster than the velocity of the low momentum pions. As already
mentioned this is
actually what one should expect, because the relevant part
of the fireball, which provides the potential, consists of pions. The
average velocity of these pions, however, is much larger then the one of the
soft pions. Therefore, as long as there are no exotic transport phenomena,
we expect the expansion velocity of the source of
the potential to be higher than the velocity of the low momentum
pions\footnote{It is actually very difficult to imagine that an attractive
two-particle interaction should slow down slow particles even more.
Let us consider a pair of pions, fast and slow, which interact by an
attractive potential. Clearly the effect of the potential
is, to slow down the fast pion and to accelerate the slow one, such
that both at the end have the same velocity.}. As a consequence the momentum
of the pion remains essentially unchanged during the expansion. Thus, as long
as
we consider only one fluid (in our case the pions) an attractive potential
does not lead to considerable enhancement in the spectrum at low momenta,
because the velocity of the particles and the expansion velocity of the
potential well are intimately related. This would be different if one had two
different
fluids, one which provides the potential, and one, the spectrum of which we
want
to study. In this case one could very well imagine that the expansion proceeds
essentially adiabatically so that all bound pions would leave the potential
with
minimal momentum.
Such a possibility actually exist at BEVALAC/SIS energy ($\sim 1 \, \rm GeV$)
heavy ion
collisions. There the pions feel a collective potential through the coupling to
the delta-hole channel \cite{EW88},
which looks very similar to the one we have studied here. In this case,
however,
the potential is provided by the nucleon fireball. Since the mass of the
nucleon
is considerably larger than the one of the pions, it could very well be that
the
expansion velocity of the nucleon fireball is smaller than the velocity of the
low momentum pions. Therefore, at BEVALAC/SIS energies the pion spectra
could reveal information about the long sought in medium pion dispersion
relation\footnote{Indeed pionic spectra at these energies exhibit an
enhancement a low energies similar to those discussed. This possibility
will be addressed in a separate publication \cite{XKK92}.}.\\
\ \\

\centerline{\bf Conclusions}
\ \\

In conclusion, we could show that the observed enhancement at low transverse
momentum in the spectra of pion cannot be accounted for by collective mean
fields. We have pointed out, that this result depends very much on the
time scales involved in the expansion. As long as the potential is provided by
the same kind of particles as the ones the spectra are
being studied of, we do not expect an
enhancement, because in general the expansion velocity is faster as the
velocity
of the soft particles. This would be different at BEVALAC/SIS energies where
the
source for the pion collective potential would be given by the nucleons, which,
as a result of their larger mass, would most likely expand slower than the soft
pions.

In the case where the particles initially have been distributed according to
the
momentum spectrum of proton--proton collisions, we found that the inclusion of
particle collisions leads to considerable enhancement at low
transverse momentum in comparison to the free decay, which would more or less
correspond to the simple folding of the p-p data.  This effect, however does
not
fully account for the enhancement observed in the data.

Taking into account the results of ref. \cite{BBD92} it seems that the soft
pion puzzle seems still unresolved. The data require either a
very short hadronization time ($\tau_0 \simeq 1 \, \rm fm/c$) or the absence of
mesons heavier than the pion. Both assumption certainly require a nontrivial
scenario in order to be acceptable.

\newpage

\centerline{\bf Figure captions}

\begin{figure}[h]
\caption{Mean field potential for $M_{res} = 770 \, \rm MeV$ for different
temperatures ($T = 150,\, 200,\, 250 \, \rm MeV$) (full lines) together with
fits
according to eqn. \protect\raf{eq:4} (dashed lines).}
\label{fig:pot2}
\end{figure}

\begin{figure}[h]
\caption{Mean field potential \protect\raf{eq:2} for different resonance
masses.
Full line: $M_{res} = 500 \, \rm MeV$,
Long--dashed line: $M_{res} = 770 \, \rm MeV$
Short--dashed line: $M_{res} = 1100 \, \rm MeV$.}
\label{fig:pot1}
\end{figure}

\begin{figure}[h]
\caption{Spectra for static potential based on eqn \protect\raf{eq:4} and
parameterizations of table \protect\ref{tab:1}. Free: full line; Set 1:
short--dashed line; Set 2: long--dashed line; Set 3: dashed--dotted line.}
\label{fig:static}
\end{figure}

\begin{figure}[h]
\caption{Dispersion relation for parameterizations of table
\protect\ref{tab:1}.
Labels as in figure \protect\ref{fig:static} }
\label{fig:disp}
\end{figure}

\begin{figure}[h]
\caption{Rapidity distribution for the two initialization schemes described in
the text. Full histogram: local thermal equilibrium; dashed histogram:
p-p distribution.}
\label{fig:rap}
\end{figure}

\begin{figure}[h]
\caption{Transverse momentum spectra for local equilibrium initialization:
Full line: no mean-field, no collision; short--dashed line: no mean-field,
collisions; long--dashed line: mean field (Set 1) collisions;
dashed--dotted line: mean field (Set 2) collisions.}
\label{fig:pt1}
\end{figure}

\begin{figure}[h]
\caption{Same as figure \protect\ref{fig:pt1}, but with p-p distribution in the
initial state.}
\label{fig:pt2}
\end{figure}
/\end{document}